\documentclass[twocolumn,aps,a4paper,superscriptaddress,bibnotes]{revtex4-1}
\usepackage{graphicx}
\usepackage{amssymb}
\usepackage{amsmath}
\usepackage{siunitx}
\usepackage{hyperref}
\usepackage{lineno}
\usepackage{ulem}
\usepackage{color}

\newcommand*\subtxt[1]{_{\textnormal{#1}}}
\DeclareRobustCommand\_{\ifmmode\expandafter\subtxt\else\textunderscore\fi} 
\newcommand{\Gstate}{$\left| {\rm b} \right\rangle$}
\newcommand{\Estate}{$\left| {\rm a} \right\rangle$}
\newcommand{\Mstate}{$\left| {\rm c} \right\rangle$}
\newcommand{\EtoGtransition}{$\left|{\rm a}\right\rangle\leftrightarrow\left|{\rm b}\right\rangle$}
\newcommand{\EtoMtransition}{$\left|{\rm a} \right\rangle\leftrightarrow\left|{\rm c}\right\rangle$}
\newcommand{\EtoGdecayrate}{$\Gamma\_{ab}$}
\newcommand{\EtoMdecayrate}{$\Gamma\_{ac}$}
\newcommand{\EtoGlambda}{$\lambda\_{ab}$}
\newcommand{\EtoMlambda}{$\lambda\_{ac}$}
\newcommand{\EtoGdetuning}{$\Delta\_{ab}$}
\newcommand{\EtoMdetuning}{$\Delta\_{ac}$}
\newcommand{\EtoGRabifreq}{$\Omega\_{ab}$}
\newcommand{\EtoMRabifreq}{$\Omega\_{ac}$}

\begin{document}

\title{Room-temperature amplified transduction of infrared to visible photons}
\author{Gibeom Son}
\affiliation{Department of Physics and Astronomy \& Institute of Applied Physics, Seoul National University, Seoul 08826, Korea}
\author{Songky Moon}
\email{Correspondence to: ygmalkuth@gmail.com}
\affiliation{Faculty of Liberal Education, Seoul National University, Seoul 08826, Korea}
\author{Seunghoon Oh}
\affiliation{SKKU Advanced Institute of Nano Technology, Sungkyunkwan University, Suwon, Korea}
\author{Junseo Ha}
\affiliation{Department of Physics and Astronomy \& Institute of Applied Physics, Seoul National University, Seoul 08826, Korea}
\author{Kyungwon An}
\affiliation{Department of Physics and Astronomy \& Institute of Applied Physics, Seoul National University, Seoul 08826, Korea}
\date{\today}

\begin{abstract}
Frequency transduction, which converts photons from one energy level to another, provides a way to bridge different quantum devices. The frequency transduction has been studied across various systems and frequency ranges, depending on the applications. 
In particular, infrared photons are ideal for long-distance communication, but their detection efficiency is often low. 
Converting infrared photons to visible light, where affordable detectors with high quantum efficiency are widely available, would offer significant advantages.
Here, we report an experimental demonstration of transduction of 1500-nm photons to 553-nm photons at room temperature using barium atoms of a three-level $\Lambda$ system. In our experiment conducted in free space, we could amplify the visible photons, achieving an internal efficiency of 1.49, exceeding unity. 
We also observed that the minimum transduction bandwidth is determined by the total decay rate of the excited state in the $\Lambda$-type energy levels. 
Moreover, we propose ways to improve the internal efficiency by 200-fold and to implement polarization-sensitive transduction in our scheme to be applicable in quantum information.
The present work is a step forward for the integration of quantum devices at different energy levels as well as for the development of efficient infrared-photon detectors.
\end{abstract}

\date{\today}
\maketitle

\section{INTRODUCTION}
In recent years, the advancement of quantum information technologies has been aided by the use of photons. Photons, by their very nature, can encode quantum information as flying qubits through polarization, optical path \cite{Wang2016_quantum_photonic_interconnect, Ciampini2016_path_polarization_states}, time-bin \cite{Zhong2015_time_energy_QKD}, or orbital angular momentum \cite{Zhou2016_OAM_quantum_interface}. Furthermore, photons can be utilized in quantum interfaces by leveraging their broad range of frequencies. For example, photons in the visible range can be efficiently detected by commercial detectors, facilitating high-precision measurement. Infrared (IR) photons, on the other hand, are highly advantageous for long-distance communication due to their minimal transmission loss in optical fibers \cite{Yin2016_MDIQKD_404km, Boaron2018_SecureQKD_421km}. Meanwhile, microwave photons are primarily employed in controlling solid-state qubits, such as superconducting qubits or quantum dots \cite{Larsen2015_nanowire_superconducting_qubit, Veldhorst2014_quantum_dot_qubit, Kim2015_microwave_qubit_operation}. Each of these quantum systems, utilizing photons of different frequency ranges, exhibits unique functionalities. However, there is currently no single, universal quantum system capable of performing all desired quantum operations across different platforms. Therefore, recent studies have focused on integrating different quantum systems, each optimized for specific fuctions, rather than pursuing a standalone, universal quantum system \cite{Kurizki2015_hybrid_quantum_technologies}. This integration has emerged as a promising approach to realizing a universal quantum device. In this context, efficient frequency transduction among photons of varying frequencies plays a crucial role in bridging the gap between quantum systems in diffrent energies. 

The frequency transduction was first proposed and demonstrated in 1990s \cite{Kumar1990_quantum_conversion, Huang1992_quantum_conversion}, with the aim of enabling the integration of quantum devices operating at different energy levels. Since then, extensive research on frequency transduction has been conducted using various platforms, including atomic ensembles \cite{Borowka2024_Rydberg_converter, Li2021_atomic_beam_transduction, Tseng2024_diamond_QFC, Tu2022_microwave_to_optics_conversion}, optomechanical systems \cite{Andrews2014_microwave_optical_conversion, Balram2016_piezo_optomechanical_circuits, Chen2021_frequency_upconversion, Fan2016_single_photon_shifter, Hill2012_optomechanics_conversion, Jiang2020_piezo_optomechanical_transduction, Vainsencher2016_piezoelectric_optomechanics}, and nonlinear materials \cite{Rutz2017_quantum_frequency_conversion, Wang2018_ultrahigh_efficiency_conversion}. These platforms have demonstrated the ability to transduce photons across a wide spectrum of frequencies from optical range to microwave.

In particular, infrared photons are highly advantageous for long-distance communication but are difficult to detect efficiently. Converting them to the visible range, where detection is significantly easier, is therefore highly desirable. While recent research has explored frequency transduction from infrared to visible light across various systems, 
\cite{Rakher2010_quantum_transduction, Guo2016_on_chip_conversion, Strassmann2019_frequency_conversion_noise}, 
achieving high efficiency remains a major challenge.
Here, we report an experimental demonstration of frequency transduction and amplification of 1.5~$\mu$m photons, advantageous for long-distance communication, into 553~nm visible photons at room temperature, utilizing the three-level $\Lambda$-type structure of Ba-138 atoms. To overcome the low transduction efficiency, we desinged an amplification process that leverages the intrinsic properties of the atoms. We compare the number of absorbed 1.5~$\mu$m photons with the number of collected 553~nm photons, yielding an amplified internal efficiency of $1.49$. Additionally, we measure the power broadening of the transduction bandwidth, whose minimum value is the total decay rate of the excited state. This work, presenting an effective room-temperature frequency transduction system, contributes to future integration of quantum platforms at different energies.

\section{THEORETICAL MODEL}

\begin{figure*}
    \centering
    \includegraphics[width=0.9\textwidth]{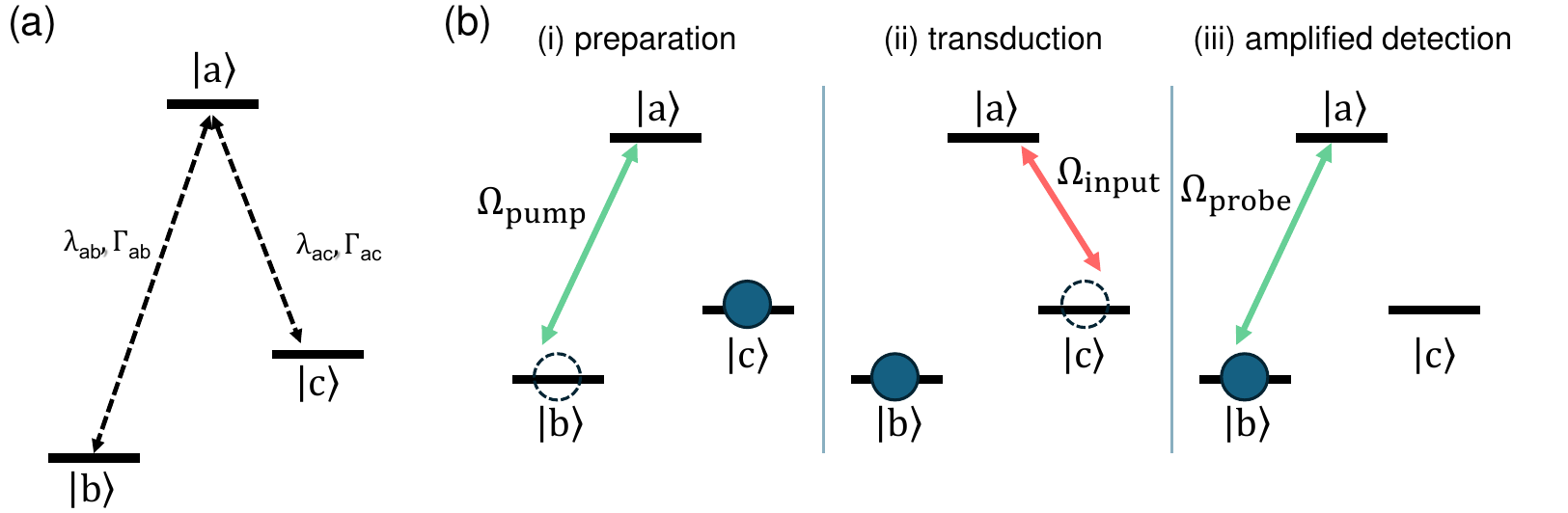}
    \caption{(a) Energy level diagram of the atom with a $\Lambda$-type three-level structure. The atom has three states \Estate, \Gstate~and \Mstate. State \Gstate~and state \Mstate~share a common excited state \Estate. Transitions \EtoGtransition~and \EtoMtransition, driven by lasers with wavelength of \EtoGlambda~and \EtoMlambda~respectively, have radiative decay rates \EtoGdecayrate~and \EtoMdecayrate~(\EtoGdecayrate$\gg$\EtoMdecayrate) associated with them, respectively.
    (b) Schematic of the transduction procedure. The atom travels through three spatially separate stages sequentially. In stage (i), the atom initially in state \Gstate~is optically pumped to state \Mstate~by driving \EtoGtransition~with a pump laser having a Rabi frequency of $\Omega\_{pump}$. In stage (ii), the atom in state \Mstate~absorbs an input photon and returns to state \Gstate. In stage (iii), the atom that has returned to state \Gstate~undergoes a cycling transition between \Gstate~and \Estate~states driven by a probe laser, emitting many photons at wavelength of \EtoGlambda.}
    \label{fig1}
\end{figure*}

We consider atoms with a $\Lambda$-type energy level structure, as depicted in Fig.~\ref{fig1}(a). The $\Lambda$-type structure is composed of two lower states -- a ground state \Gstate~and a metastable state \Mstate~-- both connected through a common excited state \Estate.  Transition \EtoGtransition (\EtoMtransition) with a radiative decay rate of \EtoGdecayrate (\EtoMdecayrate) associated with it is driven by photons at wavelength \EtoGlambda (\EtoMlambda). 
We particularly consider the case where \EtoGdecayrate$\gg$ \EtoMdecayrate~is satisfied in order to transduce photons at wavelength \EtoMlambda~into photons at wavelength of \EtoGlambda~with amplificiation.
The transduction process consists of three sequential stages: (i) preparation, (ii) transduction, and (iii) amplified detection. The atomic transition at each stage is depicted in Fig.~\ref{fig1}(b).
The Hamiltonian for each stage is given by
\begin{equation}
\begin{split}
    H =& \Delta\_{ab}\left|\mathrm{b}\right\rangle\left\langle\mathrm{b}\right| + \Delta\_{ac}\left|\mathrm{c}\right\rangle\left\langle\mathrm{c}\right| + \frac{\Omega\_{ab}}{2}\left(\left|\mathrm{b}\right\rangle\left\langle\mathrm{a}\right|+\left|\mathrm{a}\right\rangle\left\langle\mathrm{b}\right|\right) \\
    & + \frac{\Omega\_{ac}}{2}\left(\left|\mathrm{c}\right\rangle\left\langle\mathrm{a}\right|+\left|\mathrm{a}\right\rangle\left\langle\mathrm{c}\right|\right)
    \end{split}
    \label{eq.1}
\end{equation}
where \EtoGdetuning~is the detunings between the pump(probe) laser and the \EtoGtransition~transition whereas \EtoMdetuning~is the detuning between the input laser and the \EtoMtransition~transition. \EtoGRabifreq~and \EtoMRabifreq~are the Rabi frequencies of the pump(or probe) and the input lasers, respectively. In (i)-preparation stage and (iii)-detection stage, \EtoMdetuning~and \EtoMRabifreq~are 0, while in (ii)-transduction stage, \EtoGdetuning~and \EtoGRabifreq~are 0. 
The Hamiltonian in Eq.~(\ref{eq.1}) leads to the following optical Bloch equations:
\begin{equation}
    \begin{aligned}
    \dot{\rho}_{aa} &= -(\Gamma\_{ab} + \Gamma\_{ac}) \rho\_{aa} - i \frac{\Omega\_{ab}}{2} (\rho\_{ab} - \rho\_{ba}) \\
    &~~~ - i \frac{\Omega\_{ac}}{2} (\rho\_{ac} - \rho\_{ca}) \\
    \dot{\rho}_{bb} &= \Gamma\_{ab} \rho\_{aa} + i \frac{\Omega\_{ab}}{2} (\rho\_{ab} - \rho\_{ba}) \\
    \dot{\rho}_{cc} &= \Gamma\_{ac} \rho\_{aa} + i \frac{\Omega\_{ac}}{2} (\rho\_{ac} - \rho\_{ca}) \\
    \dot{\rho}_{ab} &= \left( -\frac{\Gamma\_{ab} + \Gamma\_{ac}}{2} + i \Delta\_{ab} \right) \rho\_{ab} + i \frac{\Omega\_{ab}}{2} (\rho\_{bb} - \rho\_{aa}) \\
    \dot{\rho}_{ac} &= \left( -\frac{\Gamma\_{ab} + \Gamma\_{ac}}{2} + i \Delta\_{ac} \right) \rho\_{ac} + i \frac{\Omega\_{ac}}{2} (\rho\_{cc} - \rho\_{aa}) \\
    \dot{\rho}_{bc} &= i(\Delta\_{ac} - \Delta\_{ab}) \rho\_{bc} + i \frac{\Omega\_{ab}}{2} \rho\_{ac} - i \frac{\Omega\_{ac}}{2} \rho\_{ab}
    \end{aligned}
 \label{eq.2}
\end{equation}

The atoms initially in state \Gstate~cannot absorb the input photons at \EtoMlambda. Therefore, it is necessary to prepare the atoms initially in state \Mstate, which is done in the preparation stage. By applying a strong pump laser driving \EtoGtransition~transition, the atoms are excited to state \Estate. State \Estate~has two possible decay channels: it can either decay back to state \Gstate~or to state \Mstate. The goal of the preparation stage is to pump all of the atoms to state \Mstate. Increasing the intensity of the pump laser can increase the excitation from state \Gstate~to state \Estate, but this process is limited by saturation.

\begin{figure*}
    \centering
    \includegraphics[width=0.9\textwidth]{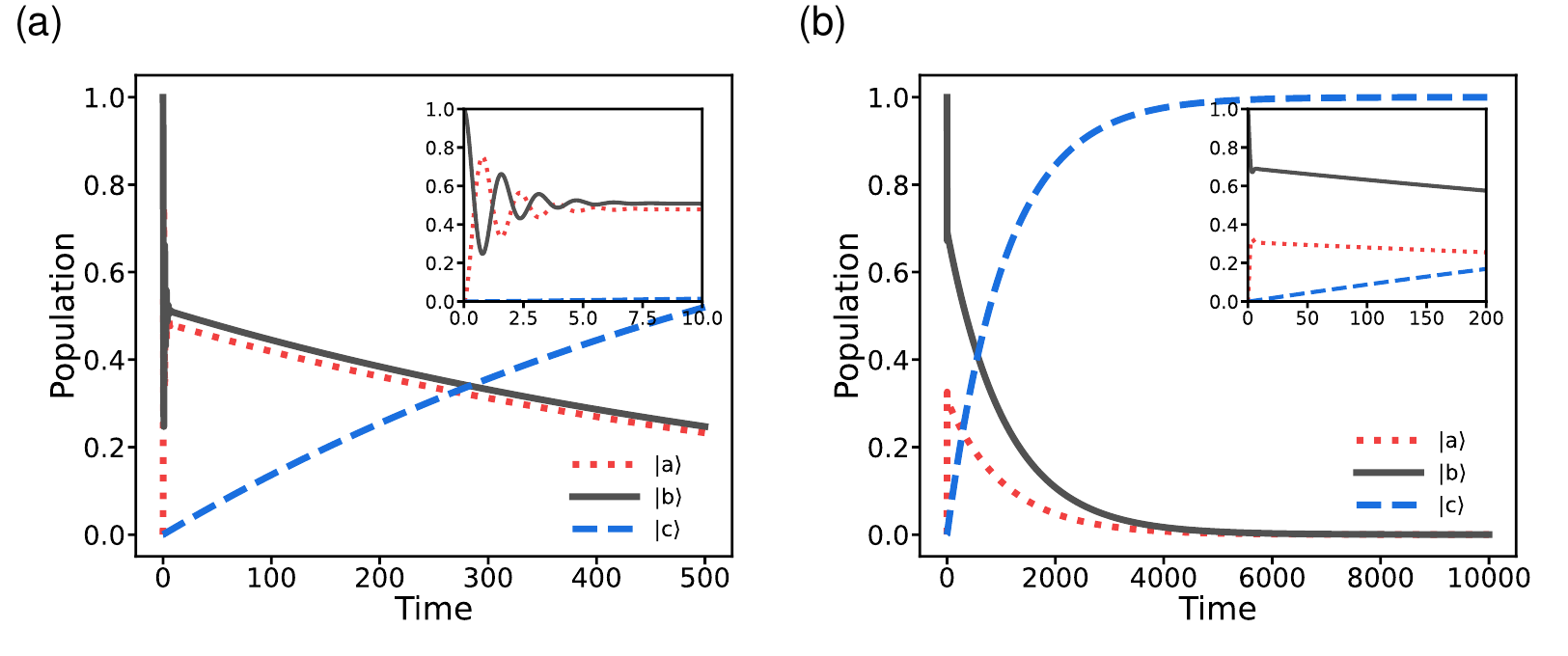}
    \caption{Time evolution of the state populations in the preparation stage. The red dotted line, black solid line, and blue dashed line represent the population of state a, state b and state c, respectively.
    (a) Time evolution of the state populations with \EtoGRabifreq = 4\EtoGdecayrate(\EtoMRabifreq=0) and interaction time of $500 \Gamma\_{ab}^{-1}$.
    (b) Time evolution of the state populations when the waist of pump beam is expanded by a factor of 20 along the atomic beam direction while keeping the total power remained the same. As a result, the Rabi frequency decreases in proportion to the square root of the area. Insets in (a) and (b) show the early-time behavior of populations. Simulation condition is as follows: \EtoGdecayrate/\EtoMdecayrate = 330, which is the case for barium-138 used in the experiment.}    
    \label{fig2}
\end{figure*}

Figure \ref{fig2} shows the result of numerical simulations based on Eq.~(\ref{eq.2}). Time evolution of the population in each state are shown in (a) and (b). In Fig.~\ref{fig2}(a), the Rabi frequency of the pump laser is set to \EtoGRabifreq = 4\EtoGdecayrate (\EtoMRabifreq~=~0). Due to the strong pump, the population in state \Estate~saturates rapidly, but because the decay rate to state \Mstate~is much smaller, most of the interaction occurs in the \EtoGtransition~transition, and not all atoms are successfully transferred to state \Mstate~for a given interaction time. Rather, the atomic beam must be exposed to the pump laser for a sufficiently long period of interaction time $T\gg \Gamma\_{ac}^{-1}$. 

Figure \ref{fig2}(b) shows the time evolution of the population when the interaction time is extended by a factor of 20 and the intensity is reduced by the same factor. Since we will perform experiments with an atomic beam, this condition is achieved by expanding the pump laser along the atomic beam direction by 20 times while keeping the total power remained the same. In this case, the excitation to state \Estate~does not reach saturation due to the reduced intensity, but the increased interaction time allows for the atoms to be successfully transferred to state \Mstate. 

Once the atoms are prepared in state \Mstate, they are capable of absorbing input photons at \EtoMlambda. At stage (ii), when an atom in state \Mstate~absorbs an input photon, it is excited to state \Estate, and from there, it decays almost exclusively to state \Gstate~with a probability close to unity. Thus, by measuring the population in state \Gstate~after interaction with the input laser, we can determine the number of input photons absorbed per atom.

After the atoms absorb input photons and return to state \Gstate, they are probed by a probe beam at stage (iii). The probe beam interacts with the atoms in state \Gstate, driving a cycling transition \EtoGtransition, which leads to the emission of amplified fluorescence at \EtoGlambda. When an atom initially prepared in state \Mstate~absorbs a single input photon and subsequently returns to state \Gstate, it can emit output fluorescence photons of \EtoGlambda~as much as the ratio of the decay rates \EtoGdecayrate/\EtoMdecayrate$\gg 1$~as it undergoes repeated cycling transitions with the probe laser. This indicates the theoretical maximum value of the amplified internal efficiency $\eta\_{max}$ would be \EtoGdecayrate/\EtoMdecayrate. However, in practice, achieving full amplification through the probe is challenging, similar to the difficulties encountered during the preparation stage. The actual number of the emitted photons would be proportional to the population that is returned to state \Mstate~by the probe laser, and can be expressed as $\eta\_{max} \cdot \rho\_{cc}(\tau\_{probe})$, where $\rho\_{cc}(\tau\_{probe})$ represents the fraction of the population that has decayed to state \Mstate~after interaction with the probe laser at \EtoGlambda.

\section{EXPERIMENTAL RESULTS}

\begin{figure*}
    \centering
    \includegraphics[width=0.85\textwidth]{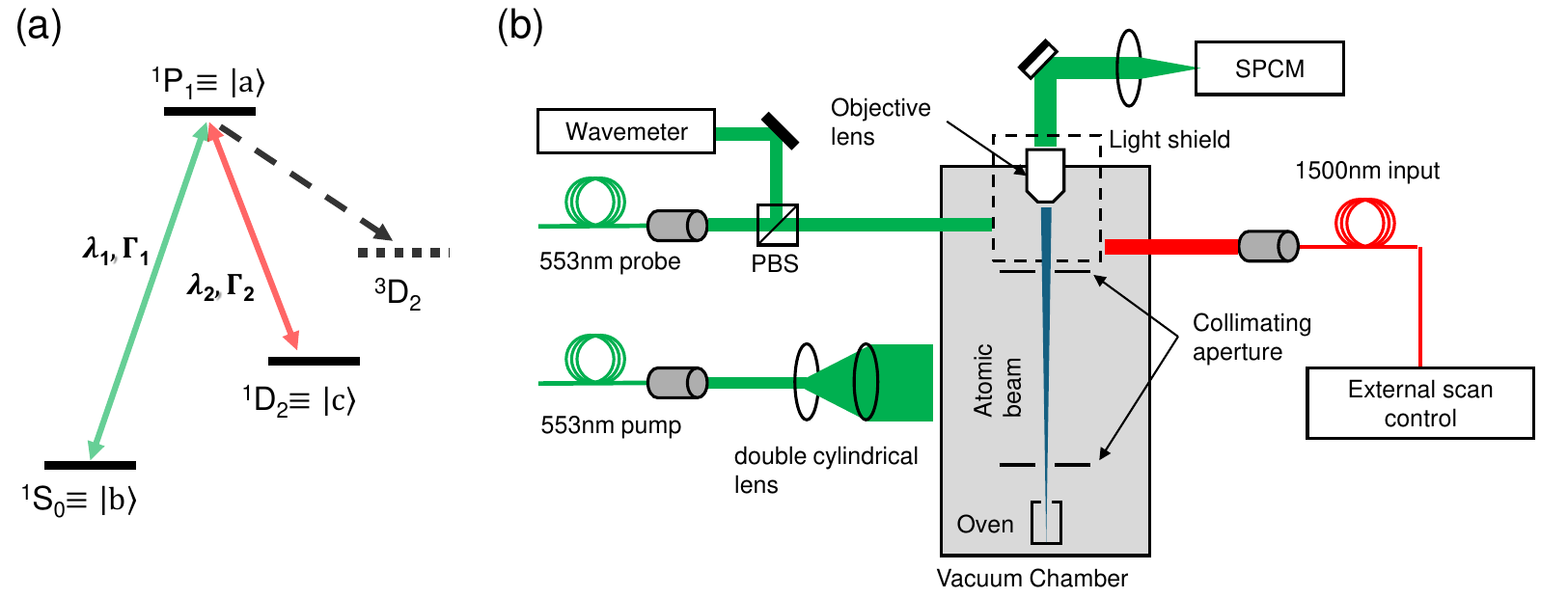}
    \caption{
    (a) Energy level diagram of the three-level lambda system of $^{138}$Ba. The $^{138}$Ba atom has three relevant energy levels $^1\mathrm{S}_0$, $^1\mathrm{P}_1$, and $^1\mathrm{D}_2$. The ground state $^1\mathrm{S}_0$ and the metastable state $^1\mathrm{D}_2$ share the excited state $^1\mathrm{P}_1$. The $^{1}\mathrm{S}_0\leftrightarrow{^{1}\mathrm{P}_1}$ and $^{1}\mathrm{D}_2\leftrightarrow{^{1}\mathrm{P}_1}$ transitions, driven by 553~nm and 1.5~$\mu$m lasers, respectively, have linewidths of $\Gamma\_{ab}/2\pi = 18.9$~MHz and $\Gamma\_{ac}/2\pi = 40$~kHz. The $^1\mathrm{D}_2$ state, with a lifetime of 0.25 s, is considered a metastable state. The $^1\mathrm{P}_1$ state can decay into various $\mathrm{D}$-states, with the $^3\mathrm{D}_2$ state being notable, exhibiting a linewidth of 28 kHz and a wavelength of 1130 nm. Among the $\mathrm{D}$-states, only the $^1\mathrm{D}_2$ state is used in the experiment, while the decay to the $^3\mathrm{D}_2$ state is considered only in theoretical calculations.
    (b) Schematic of the experimental setup. The atoms travel through a high-vacuum chamber in a collimated beam and encounter three spatially separated lasers in the sequence of pump, input, and probe. The fluorescence photons emitted after the atom interacts with the probe laser are collected by an objective lens and detected by a single-photon counting module (SPCM). The objective lens is surrounded by a light shield coated with low-reflectivity paint to minimize background light, allowing only the passage of the laser and atomic beams.}
    \label{fig3}
\end{figure*}

\begin{figure}
    \centering
    \includegraphics[width=0.46\textwidth]{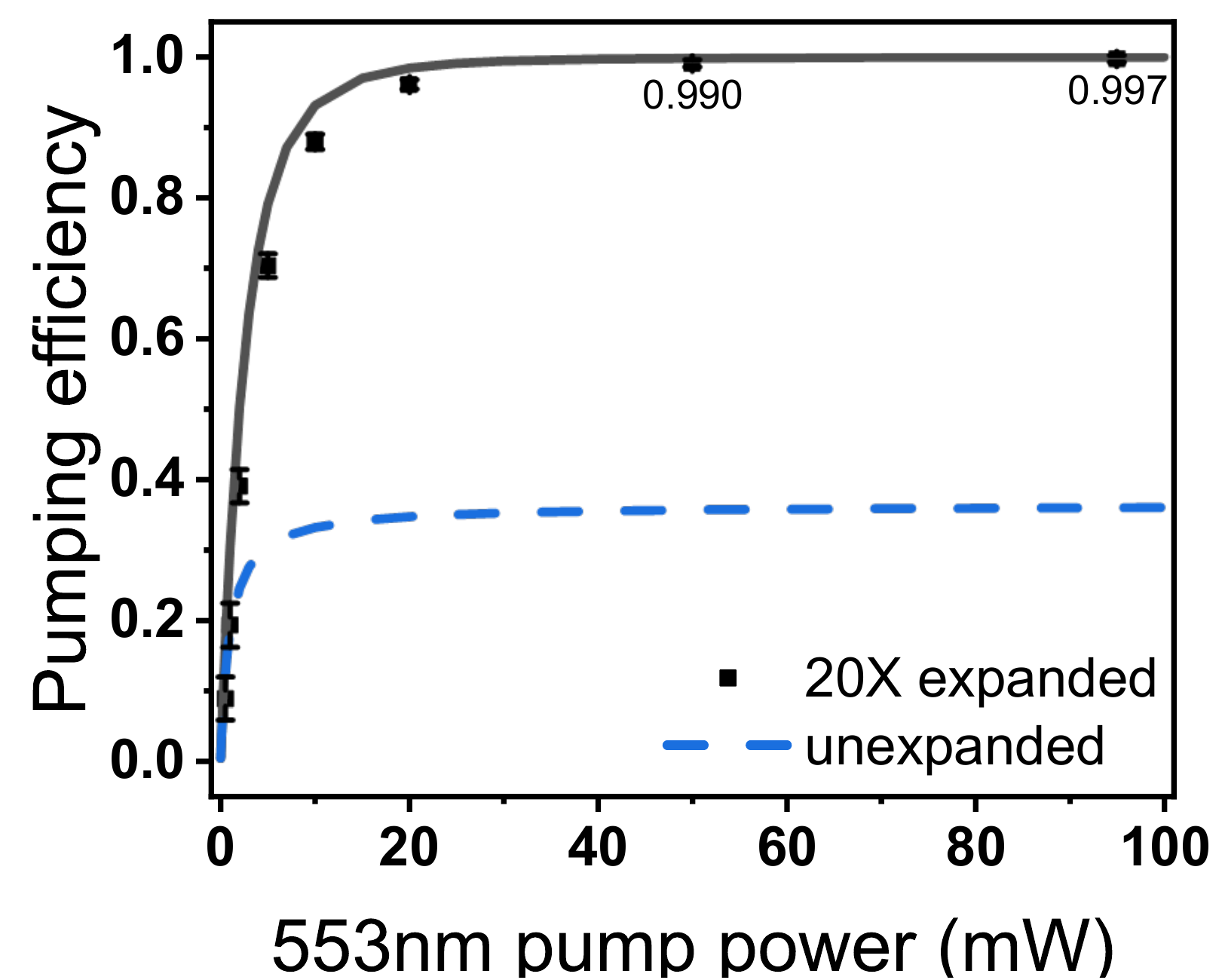}
    \caption{Plot of pumping efficiency versus pump beam power. The black squares represent data measured with the waist of pump beam expanded by a factor of 20 along the atomic beam direction as in Fig.~\ref{fig2}(b) with an interaction time of 72.4~$\mu$s. Error bars indicate standard deviation. The black solid line shows the theoretical pumping efficiency under the same conditions, while the blue dashed line indicates the expected pumping efficiency when the pump beam is not expanded.}
    \label{fig4}
\end{figure}

Barium-138 atoms form a $\Lambda$-type three level system needed in experiment. The relevant energy levels of barium is depicted in Fig.~\ref{fig3}(a). The excited state $^1\mathrm{P}_1$ corresponds to \Estate, the ground state $^1\mathrm{S}\_0$ to \Gstate~and the metastable state $^1\mathrm{D}_2$ to \Mstate.  The transition $^1$S$_0$ - $^1$P$_1$ is driven by 553~nm photons with a radiative decay rate(\EtoGdecayrate) of 18.9 MHz. Likewise, the transition $^{1}\mathrm{P}_{1}$ - $^{1}\mathrm{D}_{2}$ is driven by 1.5~$\mu$m photons with a notably smaller radiative decay rate(\EtoMdecayrate) of 40 kHz. In addition, there is a small chance for the $^1\mathrm{P}_1$ state to decay into the $^3\mathrm{D}_2$ state with a radiative decay rate of 28 kHz. 

The schematic of our experimental apparatus is shown in Fig.~\ref{fig3}(b). 
The frequency of the 553-nm laser was tuned to the $^1\mathrm{S}\_0$-$^{1}\mathrm{P}_{1}$ transition by using a wavemeter. The 553~nm pump laser was incident perpendicularly to the collimated beam of atomic barium at a distance of 28 cm from the atomic beam oven. The pump beam diameter was 2.55 mm and the atomic velocity was 750 m/s, corresponding to an interaction time of 3.62~$\mu$s. To increase the interaction time, two cylindrical lenses with focal lengths differing by a factor of 20 were used to stretch the pump beam along the atomic beam’s propagation direction. The elliptical pump beam had a reduced intensity but an extended interaction time of 72.4~$\mu$s, corresponding to 8598$\Gamma\_{ab}^{-1}$, achieving a pumping efficiency of 99.7(1)\% at 100 mW, as shown in Fig.~\ref{fig4}.

A small portion of the 553~nm laser power was split to the probe beam, which was positioned 10 mm away from the 1.5~$\mu$m input infrared laser. The probe beam intersected the atomic beam perpendicularly, and an objective lens with a numerical aperture of 0.5 used up to collect the fluorescence photons emitted by the atomic beam upon interaction with the probe beam. The collected light was then measured with a single photon counting module (SPCM).
Due to the limited field of view of the objective lens used in the experiment, the probe interaction could be observed for up to 2.92~$\mu$s.
During this time, the number of output photons that could be maximally emitted by the atom saturated by the high-intensity probe laser was 62. Given the objective lens's solid angle of 0.067, and accounting for optical losses of 0.72 and the SPCM quantum efficiency of 0.55, the number of output photons that could be detected by the SPCM was approximately 1.64. To improve the signal-to-noise ratio, the objective lens was also surrounded by a light shield coated with low-reflectance paint, allowing only the atomic beam, probe beam, and input beam to pass through. 

\begin{figure*}
    \centering
    \includegraphics[width=\textwidth]{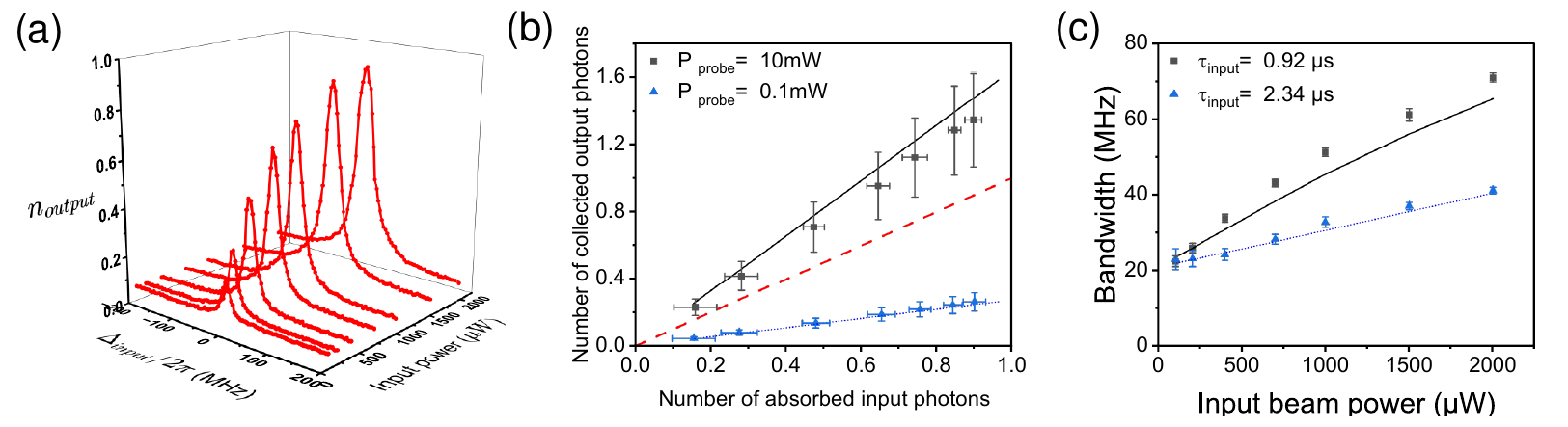}
    \caption{(a) The number of output photons as a function of the input laser detuning at various input laser power. The number of output photons is normalized by the maximum value.
    (b) The number of amplified output photons detected by the SPCM relative to the number of absorbed input photons per atom. The black squares and blue triangles represent data measured when the intensity of the probe beam was higher ($I/I\_{sat} \sim 17$) and lower ($I/I\_{sat} \sim 0.17$) than the saturation intensity, respectively. The red dashed line indicates where the efficiency equals to unity. The error bars indicate the standard deviation from repeated measurements. The black solid line and blue dotted line represents the theoretical values obtained from the solution of the optical Bloch equations without any fitting parameters.
    (c) The transduction bandwidth(FWHM) is obtained by fitting the output fluorescence versus the input laser frequency with a Lorentzian profile. The transduction bandwidth is plotted as a function of the input laser power. The black squares correspond to data measured with an interaction time of 0.92~$\mu$s between the input beam and the atomic beam, while the blue triangles represent measurements taken with an interaction time of 2.34~$\mu$s. The error bars denote the standard deviation from repeated measurements. The black solid line and blue dotted line show the fits to the data, with the Rabi frequency of the input beam used as a fitting parameter.}
    \label{fig5}
\end{figure*}

We measured the number of amplified output photons detected by the SPCM in relation to the number of input photons absorbed, as the power of the input beam was varied. Additionally, by scanning the frequency of the input beam, we measured the intensity of the output fluorescence in a fashion of excitation spectroscopy to determine the transduction bandwidth.
First, the mean number of output photons shown in Fig.~\ref{fig5}(a) represents the data obtained from the SPCM while scanning the freqeuncy of the input beam with various input beam power. At the resonance frequency, the number of output photons saturates as the input beam power increases. We found that when atoms in state \Mstate~absorb input photons and fully return to state \Gstate, the maximum number of output photons is achieved. We interpret this condition as full saturation, which allowed us to determine the number of input photons absorbed per atom.

Fig.~\ref{fig5}(b) shows the mean number of amplified fluorescence photons (output) effectively collected from the fluorescence per atom, as a function of the mean number of input photons absorbed by the same atom. The black squares represent the data when the power of the probe beam is 10 mW ($I/I\_{sat} \sim 17$), while the blue triangles represent the data obtained at a reduced probe power of 0.1 mW ($I/I\_{sat} \sim 0.17$). The number of output photons collected was determined by analyzing the fluorescence signal measured with the single photon counting module (SPCM). The count rate measured by the SPCM can be expressed by the following formula:
\begin{equation}
    \eta\_{loss} \cdot \Omega\_{solid} \cdot \frac{\Gamma\_{ab}}{\Gamma\_{ac}} \cdot \rho\_{cc}(\tau\_{probe}) \cdot \frac{nV}{\tau\_{probe}} \cdot \rho\_{bb}(\tau\_{input}),
\end{equation}
where $\eta_{\text{loss}}$ accounts for the optical losses of the detection system and the quantum efficiency of the SPCM, $\Omega\_{solid}$ denotes the solid angle captured by the objective lens relative to the total dipole radiation emitted by the atom, $\rho\_{dd}(\tau\_{probe})$ is the population in the D-state after the atom interacts with the probe beam, $n$ is the atomic density, $V$ is the volume where the atomic beam intersects the probe beam, and $\rho\_{gg}(\tau\_{input})$ is the population in the ground state after interaction with the input beam. The values used in this calculation are as follows: $n=2.82(0.59) \times 10^6$ cm$^{-3}$, $V=2.75 \times 10^{-5}$ cm$^3$, and $\tau\_{probe}=2.92~\mu\text{s}$. By dividing the measured SPCM count rate by the factor $nV/\tau\_{probe}$, we obtain the number of photons that can be collected by the SPCM from the fluorescence per atom after interaction with the probe beam.

The amplified internal efficiency, $\eta$, is defined as the ratio of the number of amplified output photons collected to the number of input photons absorbed. When the probe laser satisfies $I/I\_{sat} \approx 17$, the efficiency is measured to be 1.49. In contrast, at a lower probe power with $I/I\_{sat} \approx 0.17$, the efficiency is reduced to 0.29. The red dashed line in Fig.~\ref{fig5}(b) represents the case where the efficiency equals unity. Despite the limited interaction region observed by the objective lens due to its finite field of view, we were able to achieve an efficiency exceeding unity owing to the amplification factor $\Gamma\_{ab}/\Gamma\_{ac}$ from the cycling transition. This demonstrates that the system can be highly efficient at converting the infrared input photons into the visible output photons.

We also investigated the transduction bandwidth by using the excitation spectroscopy, where the frequency of the input beam is scanned while the output fluorescence is measured. Fig.~\ref{fig5}(c) shows the transduction bandwidth as a function of the input power. The black squares correspond to the case when the interaction time of the input beam is 0.92 $\mu\text{s}$, and the blue squares correspond to the case where the interaction time is extended to 2.34~$\mu$s. In both cases, the bandwidth converges to 21.4 MHz, as the power of the input laser decreases. This value is approximately equal to the total decay rate of the excited state.

\section{Discussion}
In Fig.~\ref{fig5}(c), we observe the trnasduction bandwidth broadens as the power of the input beam increases, exhibiting power broadening. We also note that the power broadening of the transduction bandwidth becomes more significant as the interaction time of the input beam decreases while keeping the total power the same. This dependence can be explained as follows. Since the interaction time of the input beam can be adjusted by varying its beam size, the energy transferred to a single atom during the interaction can be expressed as:
\begin{equation}
    \frac{\sigma I}{\hbar \omega} \tau\_{input} = \frac{\sigma}{\hbar \omega} \frac{P}{A} \tau\_{input} \propto \frac{\sigma}{\hbar \omega} \frac{P}{\tau\_{input}^2}\tau\_{input} \propto \frac{1}{\tau\_{input}}
\end{equation}
where $\sigma$ is the unsaturated absorption cross-section, and it is given by:
\begin{equation}
    \sigma = \sigma_0 \frac{\left( \Gamma\_{ac}/2 \right)^2}{\Delta\_{input}^2 + \left( \Gamma\_{ac}/2 \right)^2}
\end{equation}
with $\sigma_0$ the resonance value of $\sigma$, $\Delta\_{input}$ the detuning of the input beam from the resonance frequency. Thus, as the interaction time becomes shorter, the intensity of the interaction increases as $1/\tau\_{input}^2$, which allows interactions to occur even at larger detunings from resonance. This explains why the bandwidth broadens significantly with reduced interaction times.

\subsection{Improving absorption efficiency in stage (ii)} \label{sec4a}

\begin{figure*}
    \centering
    \includegraphics[width=0.7\textwidth]{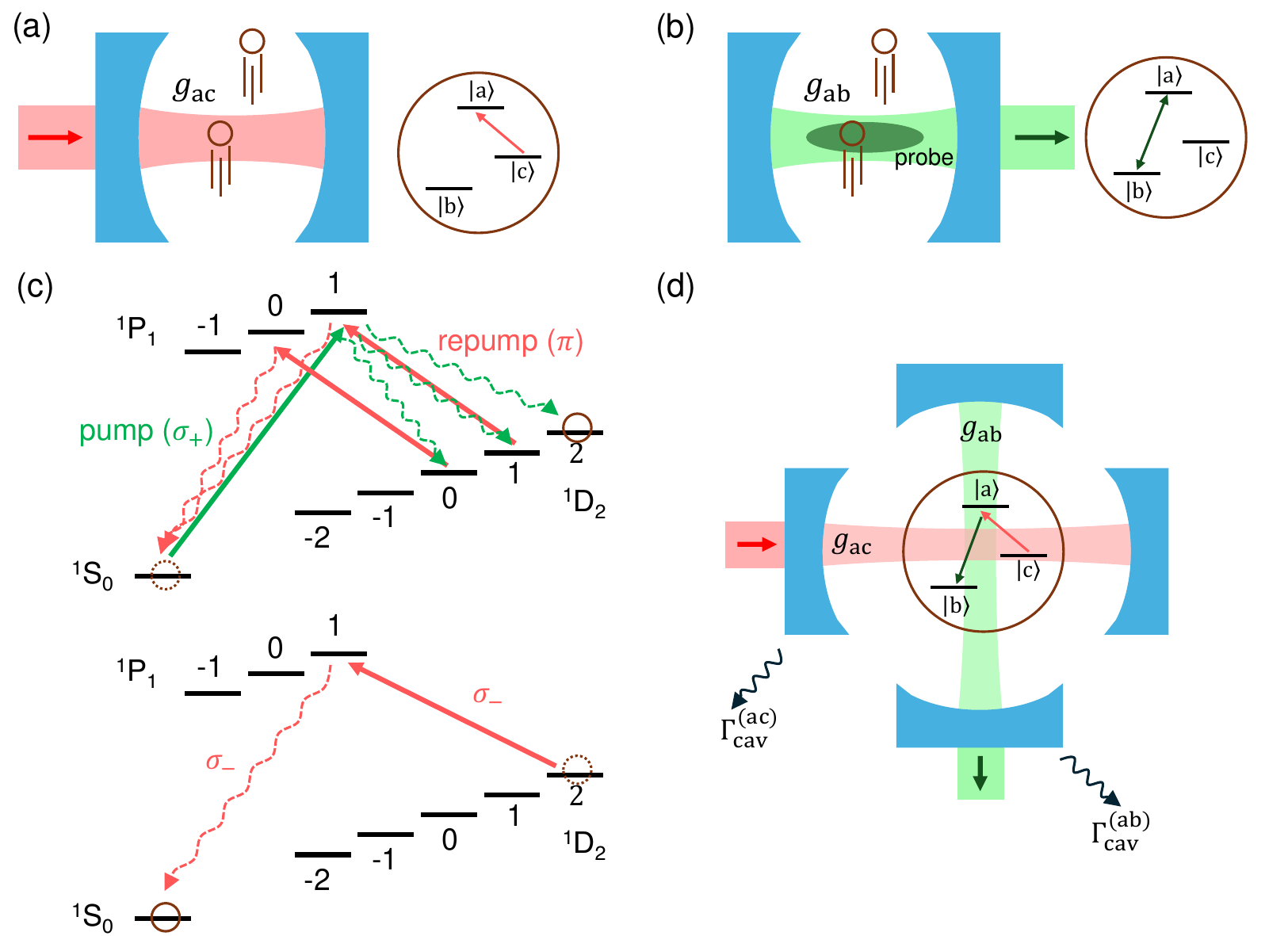}
    \caption{
        (a) Schematic of improving absorption efficiency in stage (ii). A cavity is resonant to the \EtoMtransition~transition. A coupling constant between an atom and the cavity field is given by $g\_{ac}$. Brown circles indicate atomic beam.
        (b) Schematic of improving collection efficiency in stage (iii). A cavity with decay rate of $\Gamma\_{cav}$ is resonant to the \EtoGtransition~transition. A coupling constant between an atom and the cavity field is given by $g\_{ab}$. A probe beam(dark green ellipse) is incident perpendicularly to both the cavity axis and the direction of the atomic beam.  Brown circles indicate atomic beam.
        (c) Schematic of polarization-sensitive detection. For simplicity, only $\sigma\_-$ sensitive detection is presented.
    }
    \label{fig6}
\end{figure*}

Our study highlights the usefulness of using a three-level atomic system in converting infrared photon into amplified visible photons. By exploiting the large difference in transition linewidths between the two branches in the three-level system, we achieved amplification of the converted photons, resulting in an amplified internal efficiency of 1.49.
Our experiment was conducted in free space, and thus not all input photons are absorbed by the atoms; only a fraction of the input photons interact with the atoms. The efficiency of photon absorption in this case can be characterized by the scattering ratio, $R\_{sc} = \sigma_0 / A\_{eff}$, where $A\_{eff}$ is the effective beam area. In our system, $R\_{sc}$ is in the order of $10^{-6}$. 

However, by introducing a cavity resonant to the \EtoMtransition~transition as depicted in Fig.~\ref{fig6}(a), absorption efficiency can be significantly enhanced, as this approach achieves a reduced mode volume, resulting in a high coupling constant($g\_{ac}$) and a small effective beam area simultaneously.
In a specific example using the atom-cavity coupling constant $g\_{ac}$=\EtoGdecayrate/10 and the cavity decay rate of $\Gamma\_{cav}$=\EtoGdecayrate/1000, we can achieve a near 100\% absorption efficiency according to the numerical simulations based on the master equation\cite{Scully1997_quantum_optics, Lindblad1976_quantum_dynamical_semigroups}.

\subsection{Improving collection efficiency in stage (iii)}\label{sec4b}

We use the recycling transition to amplify the output photons. Due to limited solid angle over which the photons are collected through an objective lens, only a small fraction of amplified photons are collected. 
Improvement can be made by positioning an optical cavity perpendicular to the direction of the atomic beam as depicted in Fig.~\ref{fig6}(b), thereby allowing the output photons to couple to the cavity field.
For example, if we employ the atom-cavity coupling constant $g\_{ab}=$2\EtoGdecayrate~and the cavity decay rate $\Gamma\_{cav}$=\EtoGdecayrate/2, with a probe laser with a Rabi frequency \EtoGRabifreq~= 5\EtoGdecayrate, our numerical simulations based on the master equation suggests that we can get about 800 photons decaying out the cavity, thus all collected, improving the collection efficiency by 200 times from the present 4.1 photons collected.

\subsection{Encoding quantum information}\label{sec4c}

The current scheme does not preserve the quantum information of the input photon. However, it can be modified to become polarization-sensitive and thus quantum information can be transferred from the input to the output photons.
Suppose a large Zeeman splitting is achieved as shown in Fig.~\ref{fig6}(c). In the prepation state, if we drive $\sigma\_+$ ($\sigma\_-$) transitions of $^1$S$_0\rightarrow ^1$P$_1$ with the pump laser while simultaneously performing $\pi$ pumping on $^1$D$_2\rightarrow ^1$P$_1$ transition with a repump laser, we can prepare the atoms in the $m_J = +2$ ($m_J = -2$) substate of $^1$D$_2$ state. This preparation ensures that, by selection rules, only input photons with $\sigma\_-$ ($\sigma\_+$) polarization can be absorbed, and consequency, only output photons with $\sigma\_-$($\sigma\_+$) polarization can be emitted, making the system effectively polarization-sensitive. In this way, quantum information encoded in the polarization of the input photons can be transferred to the polarization state of the output photons.

This polarization-sensitive scheme can be combined with a cavity tuned to \EtoMtransition~and another cavity tuned to \EtoGtransition~in a cross-cavity arrangement as shown in Fig.~\ref{fig6}(d). By employing $g\_{ab}=2\Gamma\_{ab}, g\_{ac}=\Gamma\_{ab}/2, \Gamma\_{cav}^{({\rm ab})}=\Gamma\_{ab}/2$ and $\Gamma\_{cav}^{({\rm ac})}=\Gamma\_{ab}/1000$, according to our numerical simulations to be published elsewhere, one can achieve near 90 \% efficiency in polarization-sensitive transduction of an single input photon to a single output photon.
Here $\Gamma\_{cav}^{({\rm ab})}$($\Gamma\_{cav}^{({\rm ac})}$) is the decay rate of the cavity tuned to \EtoGtransition (\EtoMtransition). 

\subsection{Other applications}\label{sec4d}

While our experiment utilized $^{138}$Ba atoms, our scheme is applicable to other atomic system  such as $^{88}$Sr with three-level configurations that exhibit a significant difference in linewidths between the transitions.  The significant difference in linewidths is essential not only for ensuring that the internal transduction efficiency reaches unity, allowing atoms in state \Mstate~to absorb input photons and return to state \Gstate~perfectly, but also for the amplification process of the output photons.

The ability to convert and amplify photon signals can also have practical applications in developing single-photon detectors. Single-photon detectors are essential components in quantum information technologies 
\cite{Hadfield2009_single_photon_detectors}, such as quantum key distribution \cite{Boaron2018_secure_QKD_421km, Grunenfelder2023_fast_QKD, Zhang2015_single_photon_detectors}, which ensures secure communication, and linear optical quantum computing \cite{Maring2024_quantum_computing_platform}, where precise photon detection is critical for computational accuracy. Our system, which efficiently converts infrared photons to visible photons, can be used to detect photons at specific wavelengths that correspond to atomic transition lines, offering a versatile approach to photon detection.

\section{Conclusion}
We have conducted theoretical and experimental studies on the transduction of near-infrared photons to visible photons at room temperature. Through an amplification process, we significantly increased the amplified internal efficiency, achieving the generation of more than one output photon per input photon, allowing the efficiency to exceed unity. We confirmed that the minimum transduction bandwidth is determined by the total decay rate of the excited state. 
In addition, we explored methods to enhance our results by employing the cavities and to implement polarization-sensitive transduction with near-unity efficiency for applications in quantum information.
Our results are relevant for quantum interfaces that bridge quantum devices operating at different frequencies while our high efficiency can be utilized in achieving room-temperature near-infrared single-photon detectors.

\begin{acknowledgments}
This work was supported by the Korea Research Foundation(Grant No. 2020R1A2C3009299).
\end{acknowledgments}

\bibliographystyle{naturemag}
\bibliography{references}

\end{document}